\documentclass{ws-ijmpd}
\usepackage[super,compress]{cite}
\usepackage{graphicx,array}
\usepackage{subfigure}
\usepackage{breakurl}
\usepackage{amssymb}  
\usepackage{color}
\begin{document}
\thispagestyle{plain}

\def\bib{B\kern-.05em{I}\kern-.025em{B}\kern-.08em}
\def\btex{B\kern-.05em{I}\kern-.025em{B}\kern-.08em\TeX}

\title{Towards an observational test of black hole versus naked singularity at the galactic center}

\markboth{WSPC}{Using World Scientific's \btex\/ Style File}
\author{Dipanjan Dey}
\address{International Center for Cosmology, Charusat University, Anand 388421, Gujarat, India\\
dipanjandey.adm@charusat.edu.in}
\author{Pankaj S. Joshi}
\address{International Center for Cosmology, Charusat University, Anand 388421, Gujarat, India\\
psjprovost@charusat.ac.in}
\author{Ashok Joshi}
\address{International Center for Cosmology, Charusat University, Anand 388421, Gujarat, India\\
gen.rel.joshi@gmail.com}
\author{Parth Bambhaniya}
\address{International Center for Cosmology, Charusat University, Anand 388421, Gujarat, India\\
grcollapse@gmail.com}
\maketitle
\begin{history}
\received{Day Month Year}
\revised{Day Month Year}
\end{history}
\begin{abstract} 

While the event horizon of a black hole would cast a shadow that was observed recently, a central singularity without horizon could also give rise to such a feature. This leaves us with a question on nature of the supermassive black holes at the galactic centers, and if they admit 
an event horizon necessarily.   
We point out that observations of  motion of stars around the galactic center should give a clear idea of the nature of this central supermassive object. We examine and discuss here recent developments that indicate intriguing behavior of the star motions 
that could possibly distinguish the existence or otherwise of an event horizon at the glactic center. We compare the motion of the S2 star with these theoretical results, fitting the observational data with theory, and it is seen that the star motions and precession of their orbits around the galactic center provide important clues on the nature 
of this central compact object.

\end{abstract}

\ccode{PACS numbers:} 
\section{Introduction}
Recent observations such as the Event Horizon Telescope \cite{EHT}, GRAVITY \cite{gravity,gravity1}, SINFONI \cite{SINFONI} and Galactic center group \cite{ghez,ghez2,ghez3}, have been focusing increasingly on investigating the nature of Galactic center. It is this central massive region that governs the gravitational dynamics of the entire galaxy. The centers of galaxies are very dynamic places, compacting a very large mass in a very small volume. For example, the center of our own galaxy, the Sgr A*, is as heavy as some $4-5 \times 10^ 6$  masses of the Sun, whereas the nearby galaxy M87 is Billions of the Sun masses in its center \cite{EHT}. Whenever there is a very large mass compacted in a very small region of space, astronomers generally call such a region as a `Black Hole'. Such black holes have very intriguing physical and geometrical properties and these could reveal very basic and fundamental laws for the Universe.  

The real signature of a black hole, however, is offered by the general theory of relativity. According to the Einstein theory of gravity, it is an `Event Horizon' that marks the boundary of a black hole. The event horizon encircles a region of space and time, from which nothing, not even light rays can escape away to a faraway observer in the Universe. It is a one way membrane where things can fall in but no material particles or light can escape away. 
At the center of such a black hole lies what is called a spacetime singularity. This is really a boundary of the spacetime in some sense where all the physical quantities such as matter densities, pressures and spacetime curvatures take arbitrarily large and diverging values. Thus the physics in the vicinity of such a singularity would be extreme, and possibly quantum gravity effects may dominate the physical laws here. 

General relativity (GR) predicts that such a spacetime singularity forms necessarily when large enough masses collapse under their own self-gravity \cite{OSD39}. While GR predicts occurrence of a singularity, it does not necessarily predict the formation or otherwise of an event horizon at the same time, and hence that of a black hole in the strict sense of GR. Such a singularity without an event horizon is called a Naked Singularity \cite{Joshi:1993zg,Mena:1999bz,Jhingan:1996jb,Singh:1994tb,Joshi:2011zm,Joshi:2012mk}.

Therefore, an important question regarding the nature of massive galaxy centers or the galactic nuclei is, whether the central singularity is covered by an event horizon or not. Technically, an event horizon necessarily implies the existence of a ring of unstable photons around the same, also sometimes called a photon ring or photon sphere. The strong gravitational lensing by such a photon sphere causes what is called a `shadow of the black hole', as was observed recently by the Event Horizon Telescope (EHT) \cite{EHT}.     
However, as is now known, even if the event horizon did not exist around the central singularity, a photon sphere would still form around the naked singularity for a range of physical parameters involved in the spacetime geometry \cite{Dey:2013yga, shaikh}. In such a case, it is then seen that a naked singularity also can cast a shadow, similar to the one that a black hole would create \cite{shaikh,shaikh1,shaikh2,Gyulchev:2019tvk}. 

The conclusion that follows is, we need further and more detailed theoretical as well as observational analysis, in order to decide whether the galactic central object is a black hole or a naked singularity. In other words, the existence and observation of a shadow is not sufficient, or need not ensure the existence of an event horizon necessarily in the galactic center.

Towards such a purpose, we pointed out recently that an important marker for this purpose could be the motion and behavior of stars around the galactic center \cite{parsa, Gillessen1,Mouawad}. Observationally, for past several years now, the motion of stars around galactic center is being traced quite carefully by missions such as GRAVITY, SINFONI, UCLA galactic center group, and others \cite{gravity,gravity1,SINFONI,ghez,ghez2,ghez3}. In fact, it is this observational effort that gives us an estimate of the huge mass that the galactic center contains. 
From such a perspective and from a theoretical point of view, we analyzed recently the timelike geodesics that represent the particles orbiting the galactic center, and their behavior for a black hole was  compared to a naked singularity geometry \cite{parth,parth1}. An interesting behavior that emerged is, the orbits around a naked singularity allows for a negative or opposite precession, as compared to the Schwarzschild black hole case. For negative precession, the angular distance travelled by a particle in between two successive perihelion points will be less than $2\pi$, and for positive precession, it will be greater than $2\pi$. This is an intriguing phenomenon that could potentially distinguish a black hole from a naked singularity in terms of relevant physical observables, due to their different characteristic geometries. 

However, in order to achieve the verification of such a behavior, 
and to find the nature of precession one way or the other, we need to have the observational data completed for several orbits of the stars around the galactic center, which requires more time. In the meantime, what we can do is to compare the orbits for a black hole and naked singularity geometry and one could try to find how they match with each other and the observational data, for a limited period of time. We report these partial results here, while a detailed investigation will be presented elsewhere. In the following we first briefly review the phenomenon of negative precession found for some naked singularity geometries. Further to this, the matching of orbit paths for a black hole and naked singularity geometries with current observational data are attempted. It follows that the star orbits in a black hole and naked singularity geometries match and coincide quite well for a range of parameters and time period. This implies that it is really checking the nature of the precession of the orbits that is necessary, and that is what will throw light on nature of the central object, distinguishing it in terms of a black hole or naked singularity.  

The paper is organized as follows. In section (\ref{sec2}), we discuss the basic orbit equations of Schwarzschild and JMN-1 (Joshi-Malafarina-Narayan-1) naked singularity spacetimes and briefly review the results which we got in our previous work \cite{parth,parth1}. In section (\ref{sec3}), we use those orbit equations to predict the future trajectories of S2 star which is orbiting around our galactic center Sgr A*. We fit the theoretical results with the data of astrometric positions of S2 star with a $95\%$ confidence interval. In section (\ref{sec4}),  we conclude with a discussion of the results and possible future pointers. In next section we use the units with Newton's gravitational constant $G$ and light velocity $c$ as one, but in section (\ref{sec3}), where  we fit the theoretical results with data, we consider the physical values of $G$ and $c$.

\section{Bound orbits in Spherically Symmetric and Static Spacetime}
\label{sec2}
The general form of the spherically symmetric and static spacetime is,
\begin{equation}
    ds^2 = - g_{tt}(r)dt^2 + g_{rr}(r)dr^2 + g_{\theta\theta}(r)d\theta^2 + g_{\phi\phi}(r)\sin^2\theta d\phi^2\,\, ,
    \label{static}
\end{equation}
where $g_{tt}$, $g_{rr}$, $g_{\theta\theta}$ and $g_{\phi\phi}$ are functions of $r$. The energy $(\gamma)$ per unit rest mass  and the angular momentum ($h$) per unit rest mass are conserved due to the temporal and spherical symmetries of the spacetime (\ref{static}). Using these conserved quantities and the normalization of four velocity ($u^{\mu}u_{\mu}=-1$), we can derive the general expression of effective potential, which plays an important role on the particle trajectories in the above spacetime. The general expression of the effective potential ($V_{eff}$) is given by,
\begin{eqnarray}
    V_{eff}(r) = \frac{1}{2}\left[g_{tt}(r)\left(1+\frac{h^2}{g_{\phi\phi}(r)}\right)-1\right]
    \label{veffgen}\,\, ,
\end{eqnarray}
where we consider an equatorial plane ($\theta=\frac{\pi}{2}$) for the particle trajectories. 
Using effective potential, the stable bound orbits of a freely falling particle can be described as, 
\begin{eqnarray}
   V_{eff}(r_{min})=V_{eff}(r_{max})=E\, , \,\, \nonumber\\
   E-V_{eff}(r)>0\, ,\,\,\,\forall r\in (r_{min},r_{max}).
   \label{bound}
\end{eqnarray}
where $r_{min},~r_{max}$ are the radial distances from the center to perihelion and aphelion positions of a bound orbit respectively and $E$ is the total energy of the freely falling particle. 
One can write down the equation of a orbit  of a freely falling particle using the normalization equation of the four velocity of that particle. The orbit equation for the general, spherically symmetric, static spacetime described in eq.~(\ref{static}) is, 
\begin{equation}
  \frac{d^2u}{d\phi^2}+\frac{\gamma^2g_{\phi\phi}^2(u)u^4}{2g_{tt}^2(u)g_{rr}(u)h^2}\left(\frac{dg_{tt}(u)}{du}\right)-A(u) \left(\frac{dg_{\phi\phi}(u)}{du}\right)
  +B(u) \left(\frac{dg_{rr}(u)}{du}\right)- C(u) = 0\,\, ,
  \label{orbiteqgen}
    \end{equation}
    where $u=\frac1r$ and,
$$A(u)=\left[\frac{\gamma^2g_{\phi\phi}(u)u^4}{g_{tt}(u)g_{rr}(u)h^2}-\frac{u^4}{2g_{rr}(u)}-\frac{g_{\phi\phi}(u)u^4}{g_{rr}(u)h^2}\right]\,\, ,$$
$$B(u)=\left[\frac{\gamma^2g_{\phi\phi}^2(u)u^4}{2g_{tt}(u)g_{rr}^2(u)h^2}-\frac{g_{\phi\phi}(u)u^4}{2g_{rr}^2(u)}-\frac{g_{\phi\phi}^2(u)u^4}{2g_{rr}^2(u)h^2}\right]$$
and
$$C(u)=\left[\frac{2\gamma^2g_{\phi\phi}^2(u)u^3}{g_{tt}(u)g_{rr}(u)h^2}+\frac{2g_{\phi\phi}(u)u^3}{g_{rr}(u)}+\frac{2g_{\phi\phi}^2(u)u^3}{g_{rr}(u)h^2}\right]$$
\begin{figure*}
\subfigure[Particle orbit in Schwarzschild spacetime $M_{TOT} =5$, $h=25$, $E=-0.018$]
{\includegraphics[width=61mm]{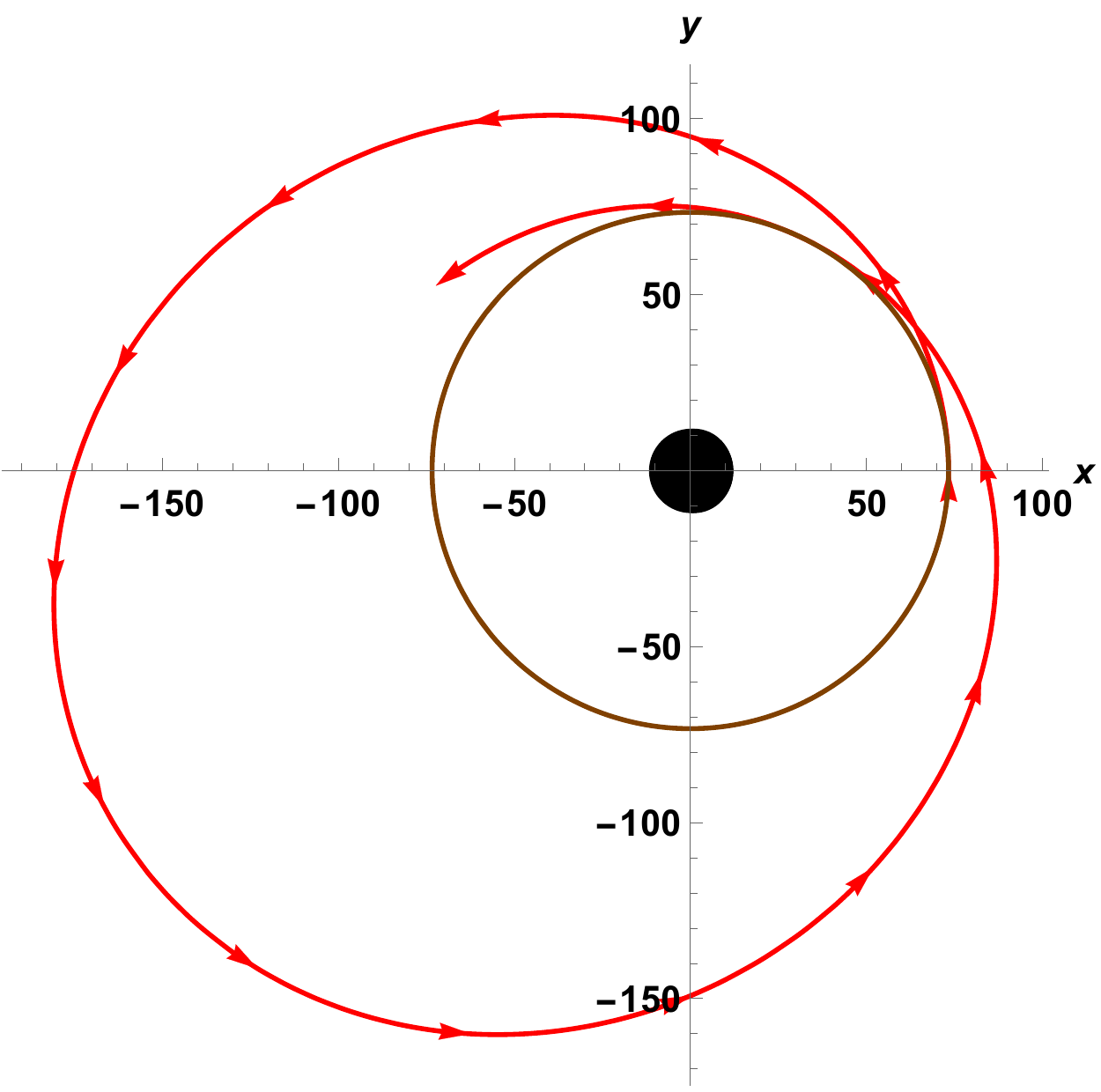}}\label{orbitsch}
\hspace{0.2cm}
\subfigure[Particle orbit in JMN-1 spacetime for $M_0 =0.5$, $~h=500$, $E=-0.2$, $R_b=1000$]{\includegraphics[width=61mm]{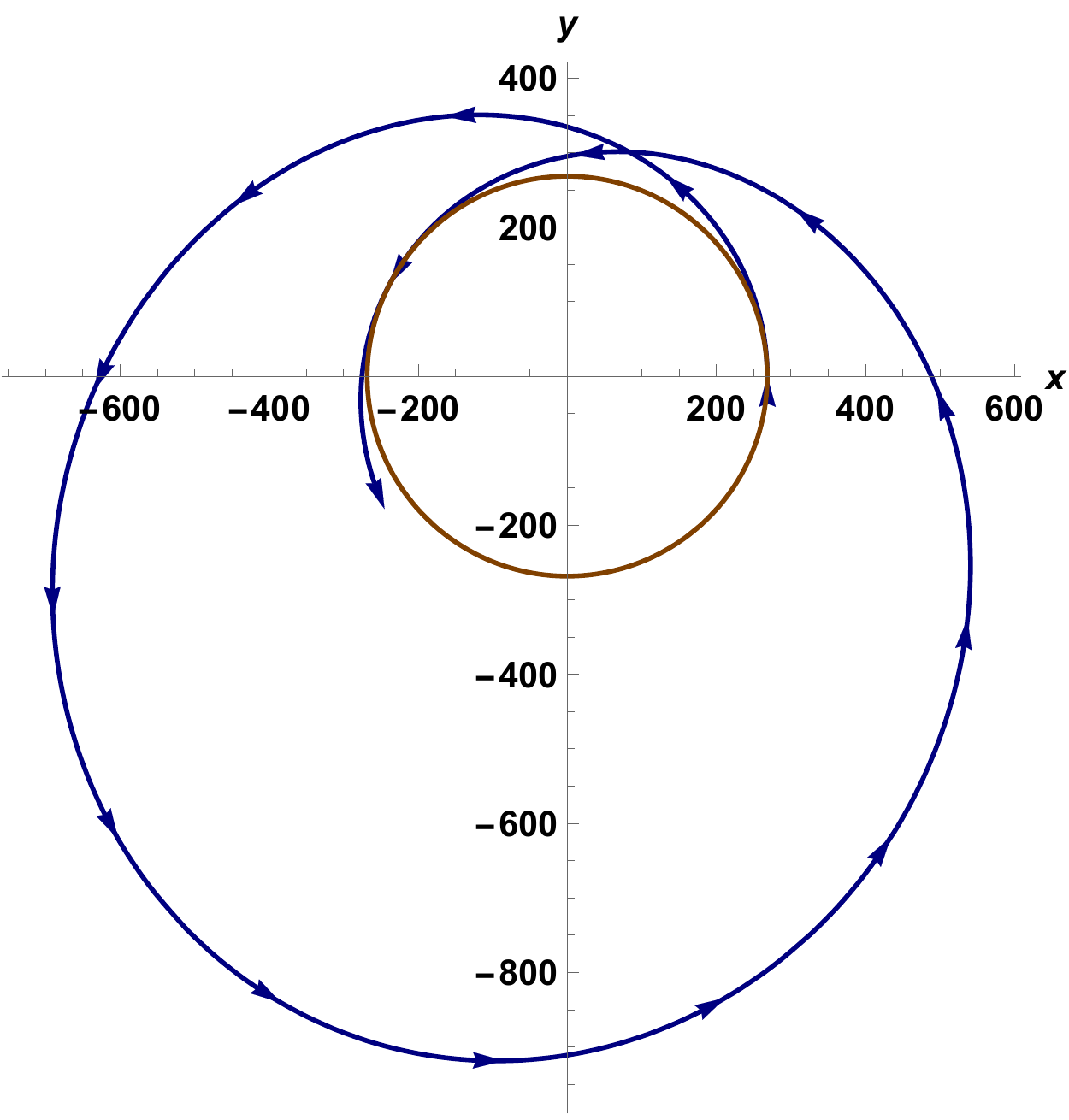}}
\label{orbitJMN1positive}
\hspace{3.2cm}
\subfigure[Particle orbit in JMN-1 for $M_0 =0.01$,$~h=25$, $E=-0.0065$, $R_b=1000$]{\includegraphics[width=65mm]{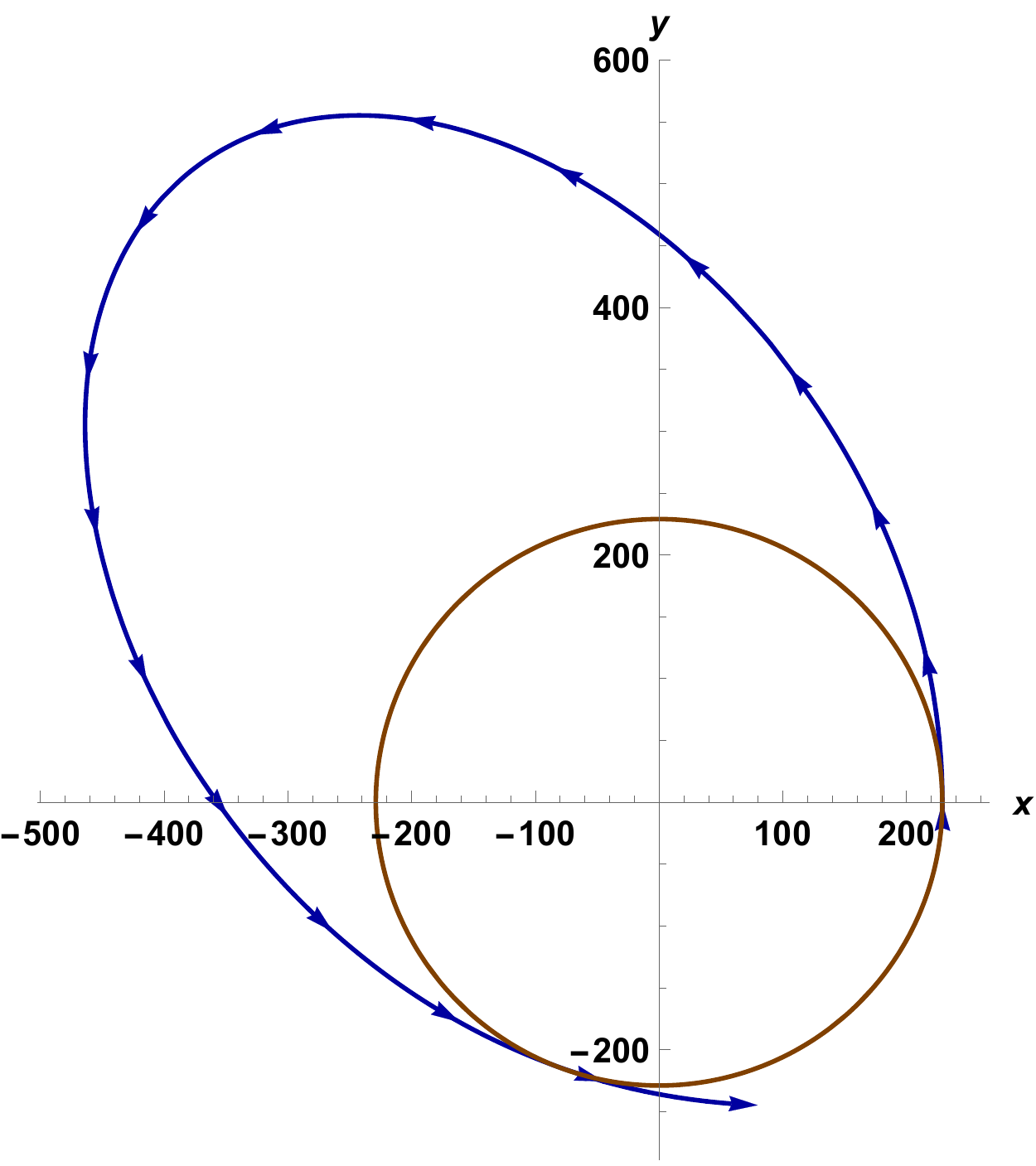}}
\label{orbitJMN1negative}
\caption{In this figure, orbits of freely falling particles in Schwarzschild and JMN-1 spacetimes are shown. It can be seen that in Schwarzschild spacetime a particle has to travel greater than $2\pi$ angular distance in between two successive perihelion points, where as in JMN-1 spacetime the angular distance can be greater or less than $2\pi$. The black dark region in the Schwarzschild spacetime shows the position of the black hole and the brown circle shows the minimum approach of the particle (perihelion points) towards the center. \label{f1}}
\end{figure*}
\begin{table}[htbp]
\centering
\subfigure{\hspace{0.2cm}Table~1. Astrometric measurements of S2 star}
\label{table:s2pos}
\begin{tabular}{l c c c c}
\hline\hline
Date & $\Delta$R.A. & $\Delta$Dec. & $\Delta$R.A. Error & $\Delta$Dec. Error \\
(Decimal) & (arcsec) & (arcsec) & (arcsec) & (arcsec) \\
\hline
2002.578  &  0.0386  &  0.0213  &  0.0066  &  0.0065 \\
2003.447 & 0.0385 & 0.0701 & 0.0009 & 0.0010 \\
2003.455 & 0.0393 & 0.0733 & 0.0012 & 0.0012 \\
2004.511 & 0.0330 & 0.1191 & 0.0010 & 0.0008 \\
2004.516 & 0.0333 & 0.1206 & 0.0009 & 0.0006 \\
2004.574 & 0.0315 & 0.1206 & 0.0009 & 0.0009 \\
2005.268 & 0.0265 & 0.1389 & 0.0007 & 0.0011 \\
2006.490 & 0.0141 & 0.1596 & 0.0065 & 0.0065 \\
2006.584 & 0.0137 & 0.1609 & 0.0033 & 0.0007 \\
2006.726 & 0.0129 & 0.1627 & 0.0033 & 0.0007 \\
2006.800 & 0.0107 & 0.1633 & 0.0033 & 0.0007 \\
2007.205 & 0.0064 & 0.1681 & 0.0004 & 0.0007 \\
2007.214 & 0.0058 & 0.1682 & 0.0004 & 0.0008 \\
2007.255 & 0.0069 & 0.1691 & 0.0010 & 0.0007 \\
2007.455 & 0.0047 & 0.1709 & 0.0004 & 0.0006 \\
2008.145 & -0.0076 & 0.1775 & 0.0007 & 0.0012 \\
2008.197 & -0.0082 & 0.1780 & 0.0007 & 0.0011 \\
2008.268 & -0.0084 & 0.1777 & 0.0006 & 0.0008 \\
2008.456 & -0.0118 & 0.1798 & 0.0006 & 0.0009 \\
2008.598 & -0.0126 & 0.1802 & 0.0009 & 0.0010 \\
2008.708 & -0.0127 & 0.1806 & 0.0008 & 0.0013 \\
2009.299 & -0.0216 & 0.1805 & 0.0006 & 0.0009 \\
2009.334 & -0.0218 & 0.1813 & 0.0006 & 0.0009 \\
2009.501 & -0.0233 & 0.1803 & 0.0005 & 0.0008 \\
2009.605 & -0.0266 & 0.1800 & 0.0012 & 0.0015 \\
2009.611 & -0.0249 & 0.1806 & 0.0006 & 0.0008 \\
2009.715 & -0.0260 & 0.1804 & 0.0006 & 0.0008 \\
2010.444 & -0.0347 & 0.1780 & 0.0013 & 0.0021 \\
2010.455 & -0.0340 & 0.1774 & 0.0008 & 0.0013 \\
2011.400 & -0.0430 & 0.1703 & 0.0009 & 0.0017 \\
2012.374 & -0.0518 & 0.1617 & 0.0012 & 0.0016 \\
2013.488 & -0.0603 & 0.1442 & 0.0006 & 0.0019 \\
2015.581 & -0.0690 & 0.1010 & 0.0014 & 0.0010 \\
\hline
\end{tabular}
\end{table}

We are going to use above eq.~(\ref{orbiteqgen}) to describe particle motion in a given spacetime. Previously \cite{parth, parth1}, we have discussed different distinguishable properties of particle orbits in black hole and naked singularity spacetimes which we  briefly review here.

As we know, the Schwarzschild spacetime is the unique static, spherically symmetric vacuum solution of Einstein field equations and it is described as,
\begin{equation}
ds^2 =  -\left(1-\frac{2M_{TOT}}{r}\right)dt^2 + \frac{dr^2}{\left(1-\frac{2M_{TOT}}{r}\right)} +r^2d\Omega^2\,\, , 
\label{schext}
\end{equation}
where the Schwarzschild radius ($R_s$) is $R_s=2M_{TOT}$. This spacetime is generally considered as the spacetime of uncharged, non-rotating black hole. 
Using eq.~(\ref{orbiteqgen}) we can derive the particle's orbits in Schwarzschild spacetime,
\begin{equation}
   \frac{d^2u}{d\phi^2} + u  =  3M_{TOT}~u^2 + \frac{M_{TOT}}{h^2}\,\, .
   \label{orbiteqsch}
\end{equation}
The first term in right hand side of the above equation is the General Relativistic correction to the Newtonian version of orbit equation.  

One can show that the final state of a continual gravitational collapse of a matter cloud depends upon the initial conditions. The final state can be a black hole or a naked singularity depending upon the initial state of the gravitational collapse \cite{Joshi:2011zm, Joshi:2013dva}. It is shown that a spherically symmetric collapsing matter cloud, with a finite pressure in it, can reach an equilibrium state at a large comoving time and the asymptotic equilibrium state is described by the spacetime which has a central naked singularity. This JMN-1 spacetime \cite{Joshi:2011zm} has the following line element,
\begin{eqnarray}
 ds^2_{JMN-1} = -(1- M_0) \left(\frac{r}{R_b}\right)^\frac{M_0}{(1- M_0)}dt^2 + \frac{dr^2}{(1 - M_0)} + r^2d\Omega^2\,\, , 
\label{JMN-1metric} 
\end{eqnarray}
and the above spacetime can be matched with an external Schwarzschild spacetime at the boundary $r=R_b$. Here the positive constant $M_0$ should be always less than 1. Using eq.~(\ref{orbiteqgen}), one can write down the following orbit equation of a freely falling particle in JMN-1
spacetime,
\begin{eqnarray}
   \frac{d^2u}{d\phi^2} + (1 - M_o) u - \frac{\gamma^2}{2h^2}\frac{M_0}{(1- M_0)}\left(\frac{1}{u}\right)\left(\frac{1}{uR_b}\right)^\frac{-M_0}{(1- M_0)}=0\,\, ,\label{orbiteqJMN-1}
\end{eqnarray}
\begin{figure}[t]
\centerline{\psfig{file=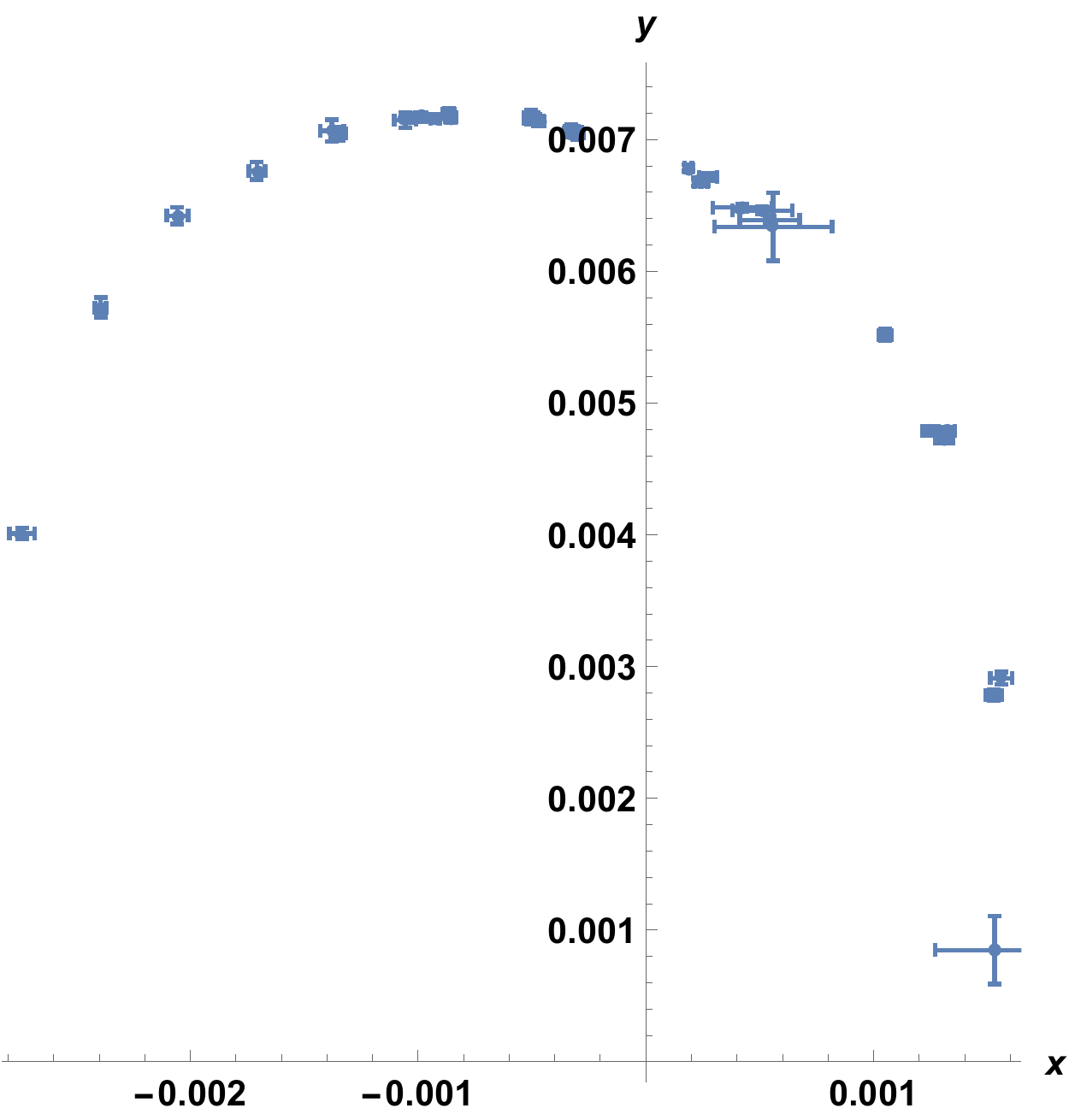,width=11cm}}
\vspace*{8pt}
\caption{Astrometric positions of S2 star at different times in (x,y) plane where the scaling unit is in parsec. \label{f1}}
\end{figure}
Using eq.~(\ref{orbiteqsch}) and eq.~(\ref{JMN-1metric}), one can show the properties of bound orbits of freely falling particles in Schwarzschild and JMN-1 spacetimes. There are some distinguishable properties of timelike orbits of paticles in JMN-1 spacetime, which cannot be seen in Schwarzschild spacetime \cite{parth}. The most important and novel feature of the bound orbits in JMN-1 naked singularity  spacetime is that the  direction of the perihelion precession can be opposite to the particle motion, which is totally forbidden in a Schwarzschild spacetime. One can show that both the negative precession (counter-precession) and positive precession (Schwarzschild like) are possible in a JMN-1 spacetime. The positive and negative precession of bound orbits in JMN-1 spacetime is shown  in fig.~(1b) and (1c) respectively, and in fig.~(1a) the positive precession in Schwarzschild spacetime is shown. It can be seen in these figures that for negative and positive precessions, the angular distance travelled by a particle in between two successive perihelion points is less and greater than $2\pi$ respectively. 

As we know, the GRAVITY collaboration is continuously observing the stellar motion around the galactic center Sgr A* \cite{gravity, gravity1,SINFONI, Alexander, ghez, ghez2, ghez3}. There are many stars (e.g. SO-2, SO-102, SO-38, etc.) orbiting around the central object Sgr A*. The perihelion points of the stars are typically $0.01-0.001$ parsec away from the galactic center. The precession of the orbits of such stars can give us the information about the nature of the central singularity. In the next section, using the data (Table 1) of the angular position of S2 star, we theoretically predict the possible different future paths of that star. 
\begin{figure*}
\subfigure[$M_{TOT} =4.63\times 10^6 M_{\odot}$,$~h=4.53101$, $~E=-2.355\times10^6$]{\includegraphics[width=61mm]{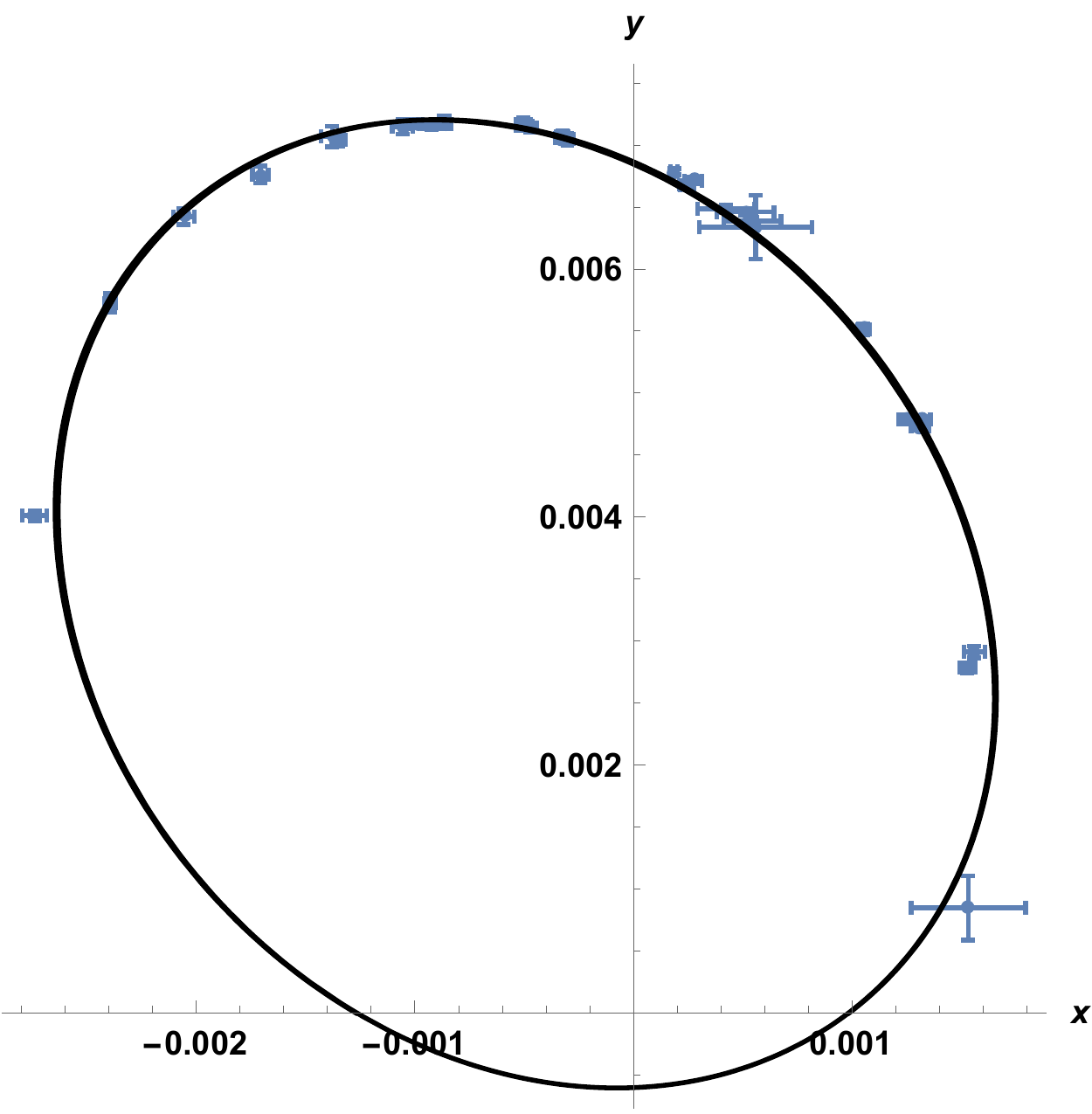}}
\label{orbitsch1}
\hspace{0.2cm}
\subfigure[$M_0=0.365, R_b=0.1~ parsec, h=385, E=-3.85\times 10^{10}$]{\includegraphics[width=62mm]{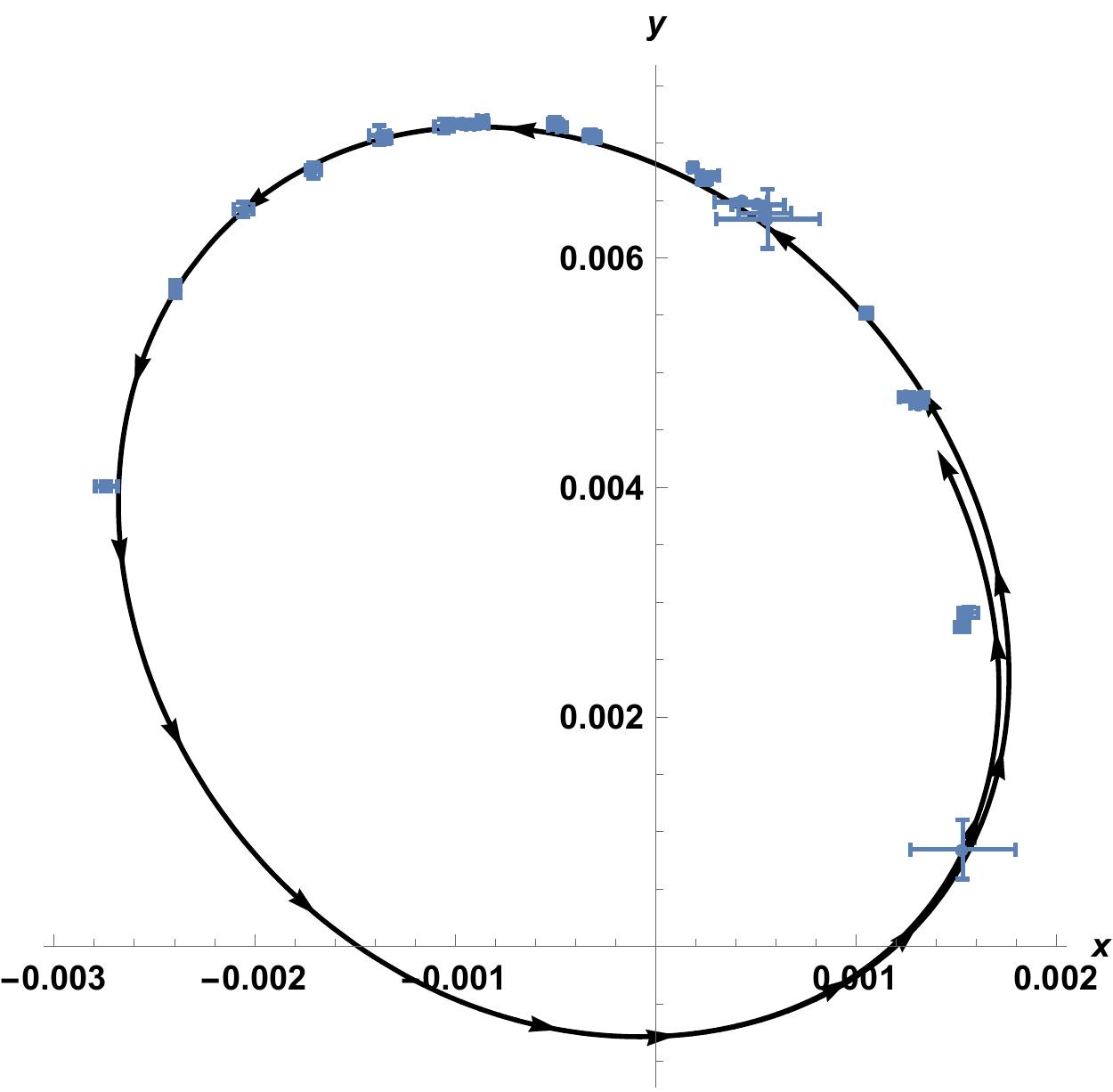}}
\label{p2}
\hspace{0.2cm}
\subfigure[$M_0=0.362, R_b=0.1~ parsec, h=385, E=-3.835\times 10^{10}$]{\includegraphics[width=61mm]{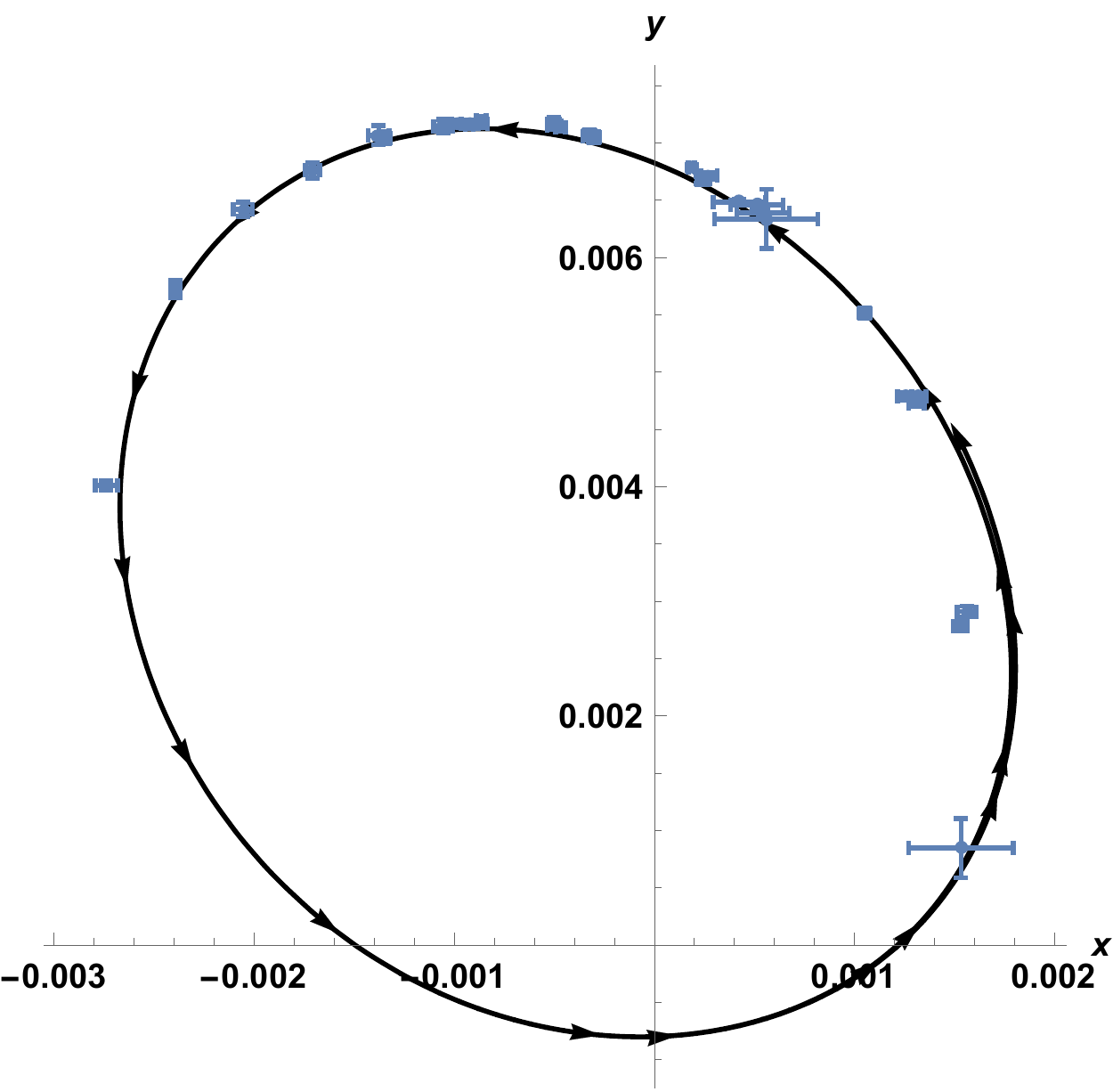}}
\label{p3}
\hspace{0.1cm}
\subfigure[$M_0=0.35, R_b=0.1~ parsec, h=383.15, E=-3,77\times 10^{10}$ ]{\includegraphics[width=60mm]{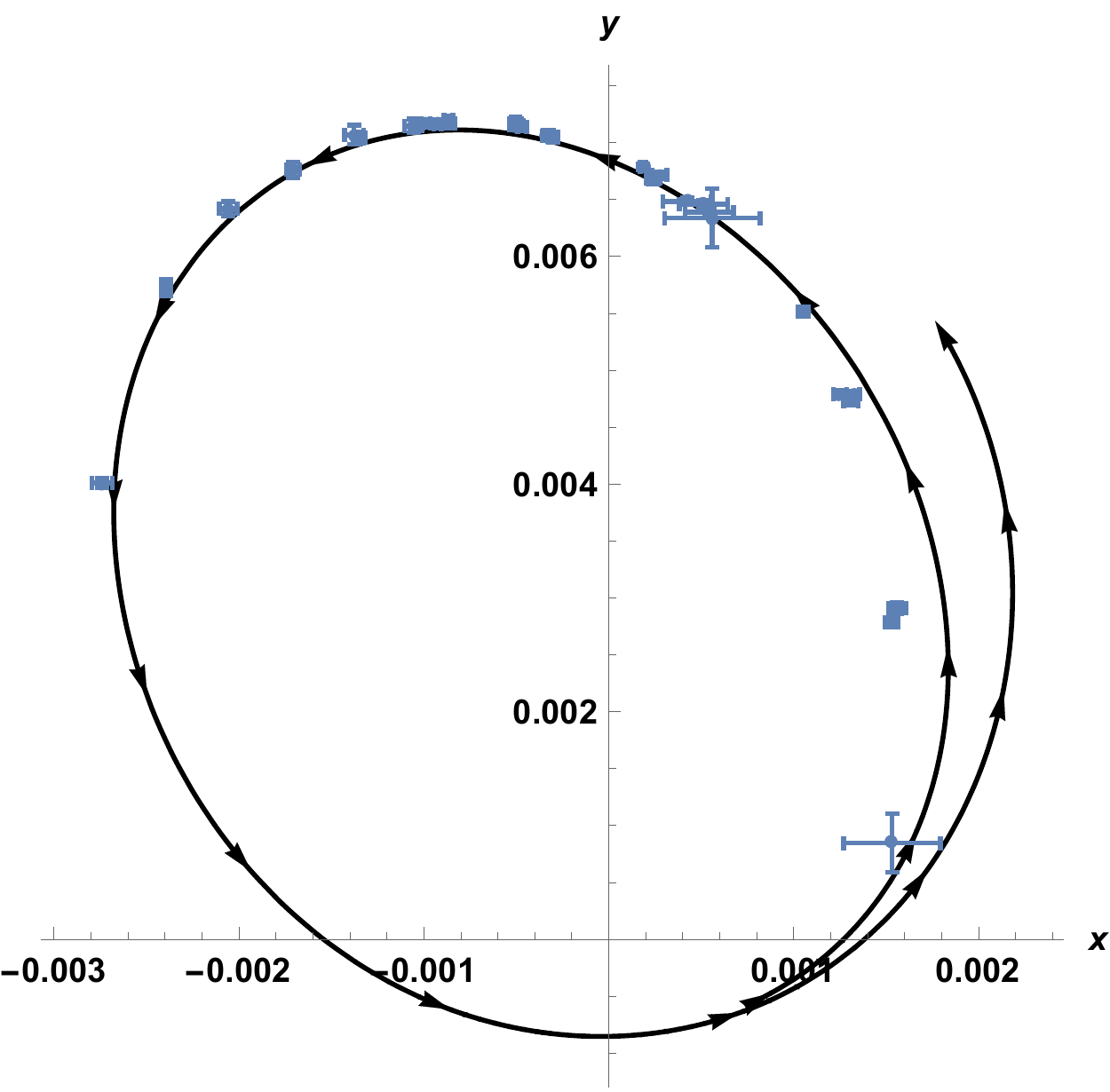}}
\label{p4}
\hspace{3.6cm}
\subfigure[$M_0=0.4, R_b=0.1~ parsec, h=415, E=-4.015\times 10^{10}$]{\includegraphics[width=61mm]{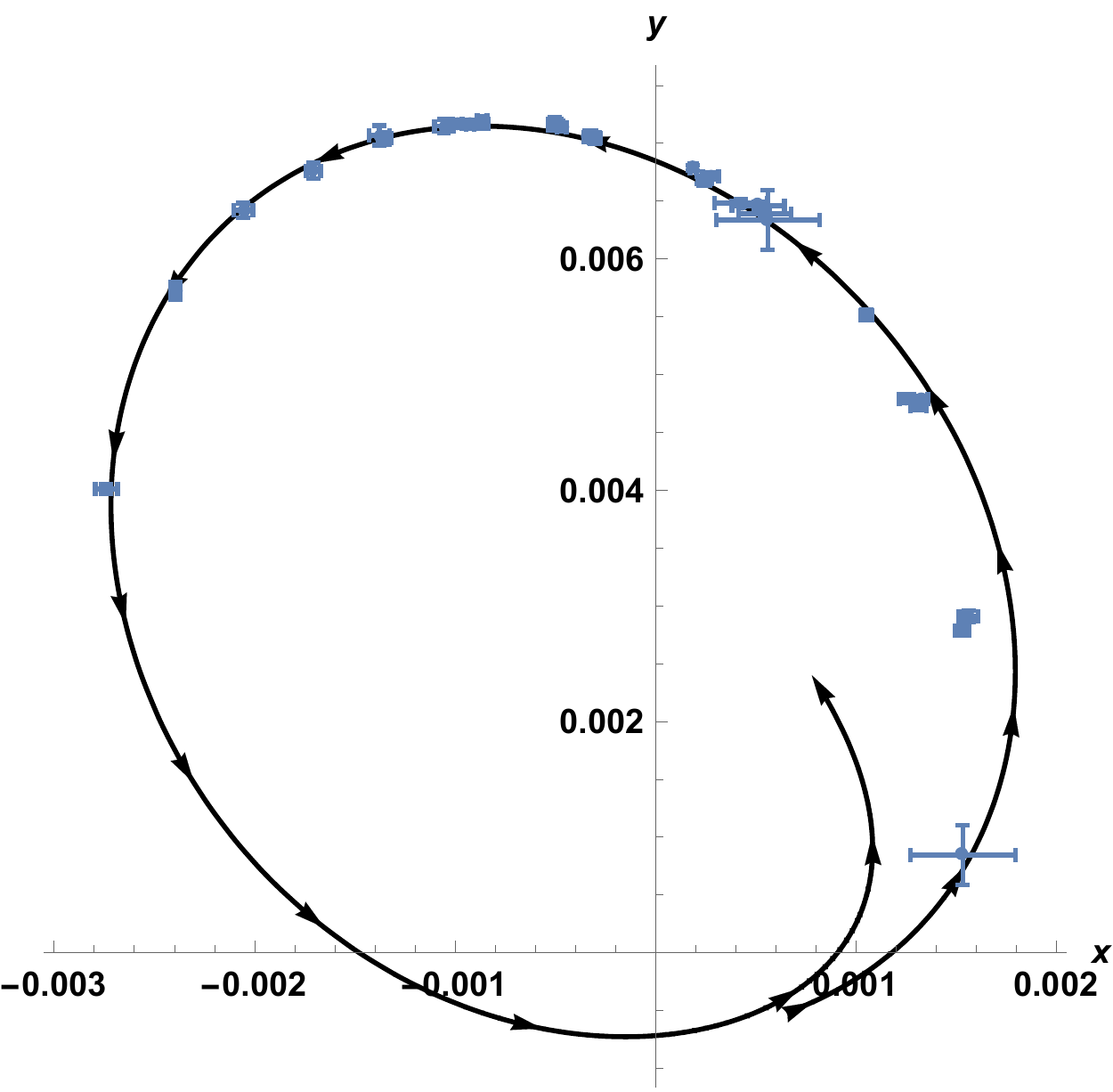}}
\label{p5}
\label{p5}
\caption{In this figure, orbit of S2 star is predicted from the best fit results with the data given in table 1. We get fig.~(3a) using the orbit equation (eq.~(\ref{SCHorbitphy})) of Schwarzschild spacetime and  rest of the other figures show the S2 star's possible motion in JMN-1 spacetime.   \label{f1}}
\end{figure*}

\section{Orbit of the S2 star around Sgr A*}
\label{sec3}
In this section, we will use physical unit (i.e. where $G$ and $c$ are not one), to discuss the orbital behaviour of S2 star. In these units, we can rewrite the orbit equations in Schwarzschild and JMN-1 spacetimes (eq.~(\ref{orbiteqsch}), (\ref{orbiteqJMN-1}) respectively) in the following form,
\begin{eqnarray}
   \frac{d^2u}{d\phi^2}&+& u - \frac{3GM_{TOT}}{c^2}u^2 - \frac{GM_{TOT}}{h^2} = 0\,\, ,\label{SCHorbitphy}\\
 \frac{d^2u}{d\phi^2}&+& (1 - M_o) u - \frac{\gamma^2}{2c^2h^2}\frac{M_0}{(1- M_0)}\left(\frac{1}{u}\right)\left(\frac{1}{uR_b}\right)^\frac{-M_0}{(1- M_0)}=0\,\, .  \label{JMN1orbitphy}
\end{eqnarray}

We now  use the above two orbit equations to compare the future trajectory of S2 star in Schwarzschild and JMN-1 spacetimes. The S2 star is orbiting around Sgr A* along an elliptical orbit with an eccentricity of 0.88 and the semi-major axis of its orbit is around 0.005 parsec. In table~(1), we have shown the astrometric measurement data of S2 star \cite{gravity, gravity1, SINFONI, Alexander, ghez, ghez2, ghez3}. In fig.~(2), we show the astrometric positions of S2 star in $(x,y)$ plane, where we transform the angular positions of that star (given in Table~(1)) into the $x$ 
and $y$ positions. In fig.~(2), the unit is in parsecs. The S2 star can follow any possible trajectory from its last known position and the future trajectory of S2 star will depend upon the causal structure of 
the central object Sgr A* of our galaxy. 

To make a theoretical prediction of the possible future trajectory of S2 star, one has to best fit the past positions of that star with the theoretical results. In fig.~(3a), we predict the future trajectory of S2 star using the eq.~(\ref{SCHorbitphy}), which is the orbit equation of a freely falling particle in the Schwarzschild spacetime. As in fig.~(3a) we use the Schwarzschild orbit equation, the orbit should have a positive precession. In this case, after best fitting the theoretical results with the observational data, we get the positive precession angle $\delta \phi\sim0.003495$ radian. For the Schwarzschild case, we fix the central Schwarzschild mass at $M_{TOT}\sim 4.3\times 10^6 M_{\odot}$. One can also use the eq.~(\ref{JMN1orbitphy}) to predict the future behaviour of the orbit of S2 star. Up to the first order approximation in eccentricity \cite{Struck:2005hi, Struck:2005oi}, one can show that the nature of precession of a bound orbit in JMN-1 spacetime depends upon the value of $M_0$ only\cite{parth,parth1}. It can be shown that with that approximation, for $M_0<\frac13$ and $M_0>\frac13$, the orbit precession will be negative and positive respectively. However, for a large eccentricity (i.e. close to one), we cannot use the approximation. For S2 star, the eccentricity is around $0.88$, therefore, first order approximation cannot give us the accurate result.  For this large value of eccentricity, one can verify that for $M_0>0.363$ and $M_0<0.363$, a positive precession and negative precession of orbit are possible respectively. Here we take $R_b=0.1$ parsec. For $M_0=0.365$, the best fit result is shown in fig.~(3b) and one can verify that the orbit has a Schwarzschild like precession with a precession angle $\delta\phi\sim 0.0035$ radians, which is very close to the precession angle calculated for the Schwarzschild spacetime. An error in the fitting of data, with $95\%$ confidence intervals can be calculated using the root mean square error (RMSE) of the data fitting. For fig.~(3b), the fitting error is $\delta x= 0.00009$ and $\delta y= 0.0002$. With a $95\%$ confidence level one can say that the data point will be inside $x\pm\delta x$ and $y\pm\delta y$, where $(x,y)$ is a point on the fitted line.  In fig.~(3c), the data fitting is done for $M_0=0.362$. With this value of $M_0$, the precession of the orbit is negative and the negative precession angle is $\delta\phi\sim -~0.0013$ radians. For this negative precession, we can get $95\%$ confidence interval with $\delta x= 0.00010257$ and $\delta y= 0.000222$ errors around the fitting line. In fig.~(\ref{f1}), the fourth (fig.~(3d)) and fifth diagram (fig.~(3e)) show large positive and negative precessions with precession angles $\delta \phi\sim 0.4605$ radians and $\delta \phi\sim -~0.128$ radians respectively. We take $M_0=0.4$ and $M_0=0.35$ for fig.~(3d) and fig.~(3e) respectively.

\section{Conclusion}
\label{sec4}

As we know, the dark matter formed the first structures in our universe and after that baryonic matter would fall into the gravitational potential well created by dark matter, and would form the galaxy like small scale structures. It can be shown that gravitational collapse of baryonic matter and dark matter, in a cosmological scenario, can terminate into an equilibrium state, and this final equilibrium can be described by some physically relevant spacetimes (e.g. the JMN spacetimes)\cite{Dey:20192,Dey:20191}. 
Therefore, the spacetimes which can be seeded by dark matter and baryonic matter should have some distinguishable signatures. There are literature where the nature of the  timelike and lightlike geodesics in different curved spacetimes are investigated \cite{Dey+15,Zhou,Levin:2009sk,levin,Chu:2017gvt,Dokuchaev:2015zxe, Borka:2012tj, Martinez:2019nor,Fujita:2009bp, Wang:2019rvq,Suzuki:1997by,Zhang:2018eau,Pugliese:2013zma,Farina:1993xw,Dasgupta:2012zf, Shoom,Eva, Eva1, Eva2, Eva3,tsirulev,Kovacs:2010xm}. 

The properties of timelike and lightlike geodesics can give us important information regarding the causal structure and dynamics of the 
central object. In this context, we showed that bound orbits in a naked singularity spacetime can have both positive and negative precession \cite{parth,parth1}. In the present paper, we have used this particular  property of naked singularity spacetimes to predict the future trajectory of S2 star. It follows that the S2 star can have negative precession if the JMN-1 naked singularity exists at the center of our galaxy. On the other hand, if S2 star shows positive precession then Schwarzschild or JMN-1 naked singularity spacetime can describe the spacetime geometry. 

There are other `S' stars which are orbiting around the central object of Sgr-A*. We can use orbit equations of different physically important naked singularity spacetimes to predict the future trajectories of those stars, along with their velocity profiles with redshift. These results will be reported separately \cite{parth3}.

\end{document}